\pdfoutput=1
\RequirePackage{ifpdf}
\ifpdf 
\documentclass[pdftex]{sigma}
\else
\documentclass{sigma}
\fi

\begin{document}

\allowdisplaybreaks

\renewcommand{\thefootnote}{$\star$}

\renewcommand{\PaperNumber}{041}

\FirstPageHeading

\ShortArticleName{Harmonic Oscillator SUSY Partners and Evolution Loops}

\ArticleName{Harmonic Oscillator SUSY Partners\\ and Evolution Loops\footnote{This
paper is a contribution to the Special Issue ``Superintegrability, Exact Solvability, and Special Functions''. The full collection is available at \href{http://www.emis.de/journals/SIGMA/SESSF2012.html}{http://www.emis.de/journals/SIGMA/SESSF2012.html}}}

\Author{David J.~FERN\'ANDEZ}

\AuthorNameForHeading{D.J.~Fern\'andez}

\Address{Departamento de F\'isica, Cinvestav, A.P. 14-740, 07000 M\'exico D.F., M\'exico}
\Email{\href{mailto:david@fis.cinvestav.mx}{david@fis.cinvestav.mx}}
\URLaddress{\url{http://usuarios.fis.cinvestav.mx/david/}}

\ArticleDates{Received May 28, 2012, in f\/inal form July 04, 2012; Published online July 11, 2012}

\Abstract{Supersymmetric quantum mechanics is a powerful tool for generating exactly solvable potentials departing from a given initial one. If applied to the harmonic oscillator, a family of Hamiltonians ruled by polynomial Heisenberg algebras is obtained. In this paper it will be shown that the SUSY partner Hamiltonians of the harmonic oscillator can produce evolution loops. The corresponding geometric phases will be as well studied.}


\Classification{81Q60; 81Q05; 81Q70}

\renewcommand{\thefootnote}{\arabic{footnote}}
\setcounter{footnote}{0}

\section{Introduction}

In the last decades there has been a growing interest in studying {\it evolution loops} (EL), which are circular dynamical processes such that the evolution operator of the system becomes the identity at a certain time \cite{fe94,fe92,fm94,fnos92,fr97,li03,li04,mi86b,mi77,mi86a,mr10,mr11}. They represent a natural generalization to what happens for the harmonic oscillator. Their importance rests on the fact that the EL are quite sensitive to external perturbations, so they are a good starting point to implement the {\it dynamical manipulation} for approximating an arbitrary unitary operator~\cite{mi86b} (see also~\cite{fe12}).

On the other hand, the {\it polynomial Heisenberg algebras} (PHA) are deformations of the Heisenberg--Weyl algebra in which the commutators of the Hamiltonian~$H$ with the annihilation~$L^-$ and creation~$L^+$ operators are standard but the commutator between the last two turns out to be a polynomial in $H$ \cite{ad94,as97,acin00,aaw99,bf11b,bf11a,cfnn04,cpr07,dek92,fh99,fnn04,ma10,mn08,mi84,vs93}. In order to characterize the spectrum of $H$, Sp($H$), one needs to determine the eigenstates of $H$ which are extremal (annihilated by~$L^-$) and have physical interpretation: thus, Sp($H$) is composed of several independent ladders (either of f\/inite or inf\/inite lengths) departing from those extremal states (for alternative deformations of the Heisenberg--Weyl algebra see e.g.~\cite{dk12,dk10,hpv10, pl97}).

In addition, nowadays it is widely accepted that {\it supersymmetric quantum mechanics} (SUSY QM) is the simplest technique for generating new Hamiltonians $H_k$ departing from a given initial one~$H_0$ (for recent books and review articles see \cite{ac04,ba01,bs04,cks01,do07,fe10,ff05,gmr11,mr04,su05}). After applying the method, it turns out that Sp($H_k$) dif\/fers little from Sp($H_0$) (the dif\/ferences rely in a f\/inite number of levels). Moreover, the algebraic structure of $H_k$ can be obtained straightforwardly of the corresponding algebra of~$H_0$. In particular, the SUSY partners of the harmonic oscillator Hamiltonian turn out to be ruled by polynomial Heisenberg algebras~\cite{fh99, mi84}: this is the simplest way for realizing in a non-trivial way such non-linear algebras.

In this paper I would like to explore in detail the possibility that the SUSY partners Hamiltonians of the harmonic oscillator can have EL, as it happens for the initial system. If the answer becomes positive, it is natural to evaluate then the geometric phase, associated to an arbitrary initial state which is cyclic \cite{aa87,acw97,be84,bbk91,cj04,fe94,fnos92,fr97, li03,li04,mo02,sw89}. It is worth to notice that some partial results along this way have been derived previously~\cite{fe94}. However, they were found just for the Abraham--Moses family of potentials isospectral to the harmonic oscillator~\cite{mi84}. Here, we are going to generalize these results for a SUSY transformation of order $k\geq 1$ so that $H_0$ and $H_k$ are not necessarily isospectral~\cite{fh99}.

The paper has been organized as follows. In the next section we shall introduce brief\/ly the evolution loops, with a discussion about the geometric phases which naturally will arise for systems satisfying such operator identity. In Section~\ref{section3} the polynomial Heisenberg algebras will be presented in general, while in Section~\ref{section4} the same shall be done for supersymmetric quantum mechanics. The SUSY partners of the harmonic oscillator will be derived at Section~\ref{section5}, including the analysis of their connections with polynomial Heisenberg algebras, evolution loops and associated geometric phases. Our conclusions shall be presented at Section~\ref{section6}.

\section{Evolution loops}\label{section2}

\begin{figure}[t]
\centering
\includegraphics[scale=0.3]{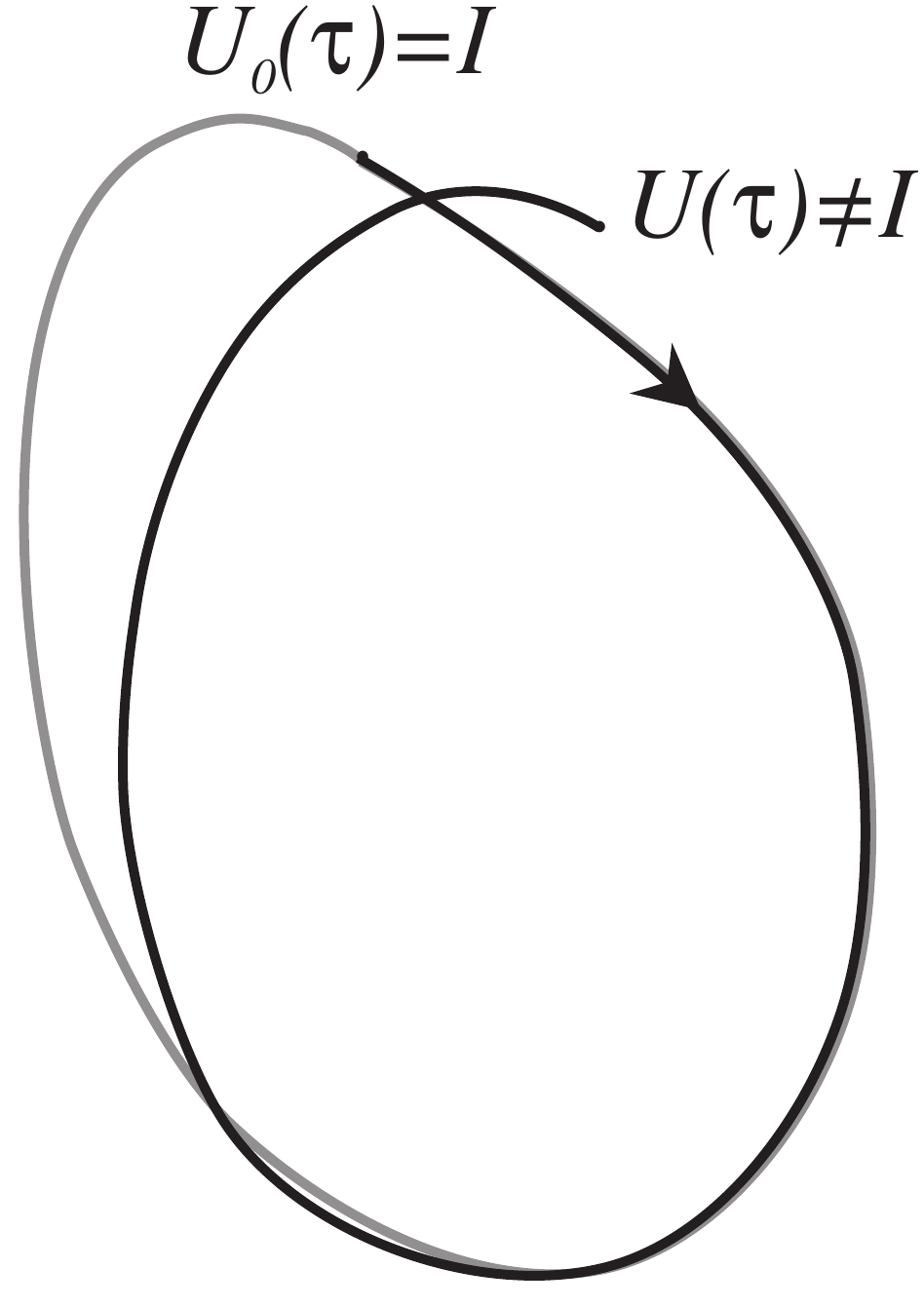}
\caption{A system which performs an evolution loop at $t=\tau$ when perturbed, in general, will deviate from this cyclic dynamical process.}\label{Fig1}
\end{figure}

At operator level, the dynamics of a quantum system is determined by its evolution opera\-tor~$U(t)$, which satisf\/ies:
\begin{gather*}
\frac{dU(t)}{dt} = -i H(t) U(t), \qquad U(0) = I,
\end{gather*}
where $H(t)$ is the system Hamiltonian, $I$ is the identity operator, $U(t)$ is unitary. As it was pointed out previously, we are interested in studying system having evolution loops, i.e., circular dynamical processes such that $U(t)$ becomes the identity (up to a phase factor) at a certain time,
\begin{gather}
U(\tau) = e^{i\varphi}I, \label{ELcondition}
\end{gather}
with $\tau>0$ being the loop period, $\varphi\in{\mathbb R}$ \cite{fe94,mi86b, mi86a}. The simplest system with an evolution loop of period $\tau=2\pi$ is the harmonic oscillator since
\begin{gather*}
U(t) = \sum_{n=0}^{\infty} e^{- i H t} \vert \psi_n\rangle\langle \psi_n\vert =
e^{- i t/2} \sum_{n=0}^{\infty} e^{- i nt} \vert \psi_n\rangle\langle \psi_n\vert \quad
\Rightarrow \\ U(\tau) = e^{- i \pi} \sum_{n=0}^{\infty} \vert \psi_n\rangle\langle \psi_n\vert =
- I,
\end{gather*}
where we have employed that $H \vert \psi_n\rangle = E_n \vert \psi_n\rangle$ with $E_n = n + 1/2$, $n=0,1,\dots$ (we are using natural units such that $\hbar=m=\omega=1$). The EL are important since they can be used as a~starting point to implement the {\it dynamical manipulation} in order to approximate any unitary operator \cite{mi86b, mi77,mi86a}. In fact, there is a prescription for implementing this kind of manipulation, which was introduced some years ago~\cite{mi86b}: f\/irst of all the system has to be placed in an~EL, i.e., $U_0(\tau)\equiv I$; by perturbing then the EL, the small deviations of this dynamical process such that $U(\tau)\neq I$ will eventually approximate an arbitrary unitary operator (see an illustration in Fig.~\ref{Fig1}).

It is worth to notice that the EL have been mainly studied for systems ruled by time-dependent Hamiltonians, either in one or several dimensions \cite{fm94,li04,mi86b,mi77,mi86a,mr10,mr11} or for purely spin systems \cite{fnos92,fr97,li03}. However, there are several works where the evolution loops are produced by time-independent Hamiltonians \cite{fe94, fe92}. In particular, an interesting physical system of such a~type consists of a charged particle inside an ideal Penning trap \cite{fe92}.

Since it is an operator relationship (compare with \cite{ezan00}), the requirement of equation (\ref{ELcondition}) is quite strong: it implies that for a system having an EL any $\vert\psi\rangle \in {\cal H}$, taken as an initial condition, becomes cyclic with period $\tau$:
\begin{gather}
\vert\psi(\tau)\rangle = U(\tau)\vert\psi\rangle = e^{i\varphi}\vert\psi\rangle.
\label{cyclicstate}
\end{gather}
Thus, it is natural to ask if the global phase $\varphi$ has a~geometric component. As it was noticed by Aharonov and Anandan~\cite{aa87}, it turns out that a geometric phase $\beta$ can be associated to any cyclic evolution $\vert\psi(t)\rangle$ satisfying equation~(\ref{cyclicstate}):
\begin{gather*}
\beta = \varphi + \int_{0}^{\tau} \langle \psi(t) \vert H(t) \vert \psi(t) \rangle  dt. 
\end{gather*}
In particular, if the system is ruled by a time-independent Hamiltonian $H(t) = H$ the previous expression becomes simpler~\cite{fe94}:
\begin{gather}
\beta = \varphi + \tau \langle \psi \vert H \vert \psi \rangle. \label{timeindependentgp}
\end{gather}

\begin{figure}[t]
\centering
\includegraphics[scale=0.45]{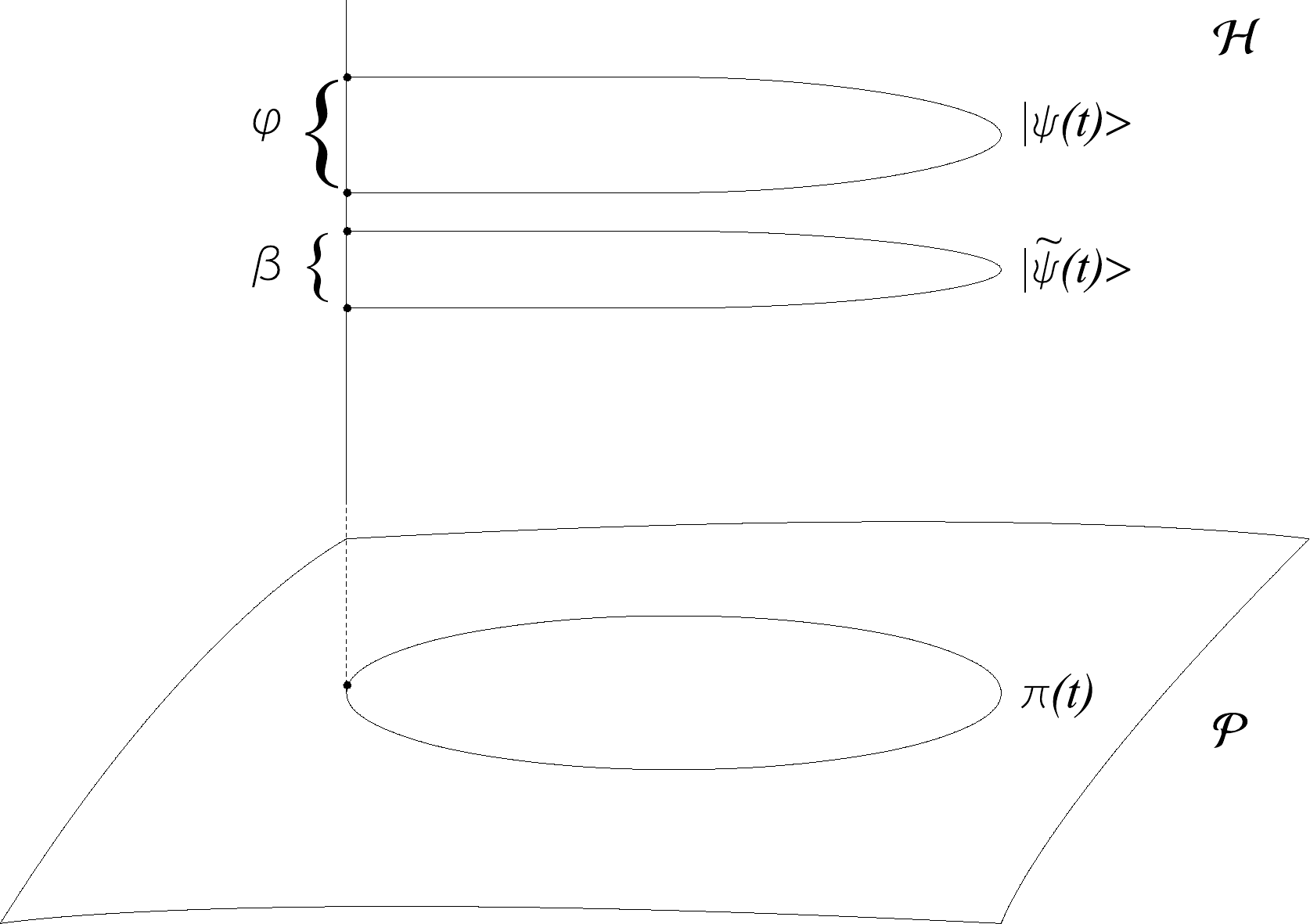}

\caption{Representation of the cyclic state $\vert \psi(t) \rangle \in {\mathcal H}$ induced by
Schr\"odinger equation, its closed {\it shadow} $\pi(t) \in {\mathcal P}$ and the corresponding
horizontal lifting $\vert \widetilde\psi(t) \rangle \in {\mathcal H}$. The holonomy of this
lifting is the geometric phase factor $e^{i\beta}$.}\label{Fig2}
\end{figure}

Let us note that $\beta$ describes a global ``curvature ef\/fect'' arising on the space of physical states of the system, which is the projective space ${\cal P}$ formed by the rays or the density operators~$\vert \psi \rangle \langle \psi \vert$ instead of the Hilbert space~${\cal H}$ \cite{aa87,bbk91,fe94,fnos92,fr97}. Due to this curvature, the horizontal lifting (parallel transport) of the closed trajectory $\pi(t)=\vert \psi(t) \rangle \langle \psi(t) \vert \in  {\cal P}$ leads to a trajectory~$\vert\widetilde\psi(t)\rangle$ which is, in general, open on ${\cal H}$. The holonomy of this lifting is the Aharonov--Anandan geometric phase factor $e^{i\beta}$ (see Fig.~\ref{Fig2}).

\section{Polynomial Heisenberg algebras}\label{section3}

The polynomial Heisenberg algebras are deformations of the Heisenberg--Weyl algebra of kind \cite{cfnn04, fh99}:
\begin{gather}
  [H,L^\pm] = \pm L^\pm, \label{pha1} \\
 [L^-,L^+] \equiv Q_{m+1}(H+1) - Q_{m+1}(H) = P_m(H), \label{pha2}
\end{gather}
where
\begin{gather}
  Q_{m+1}(H) = L^+ L^- = \prod\limits_{i=1}^{m+1} \left(H - {\cal E}_i\right) \label{pha3}
\end{gather}
is a $(m+1)$-th order polynomial in $H$ which implies that $P_{m}(H)$ is a polynomial of order $m$-th in $H$. A simple way of realizing the algebra of equations (\ref{pha1})--(\ref{pha3}) is to suppose that $H$ has the standard Schr\"odinger form,
\begin{gather*}
  H = -\frac12 \frac{d^2}{d x^2} + V(x), 
\end{gather*}
while $L^\pm$ are $(m+1)$-th order dif\/ferential operators.

\begin{figure}[t]
\centering
\includegraphics[scale=1]{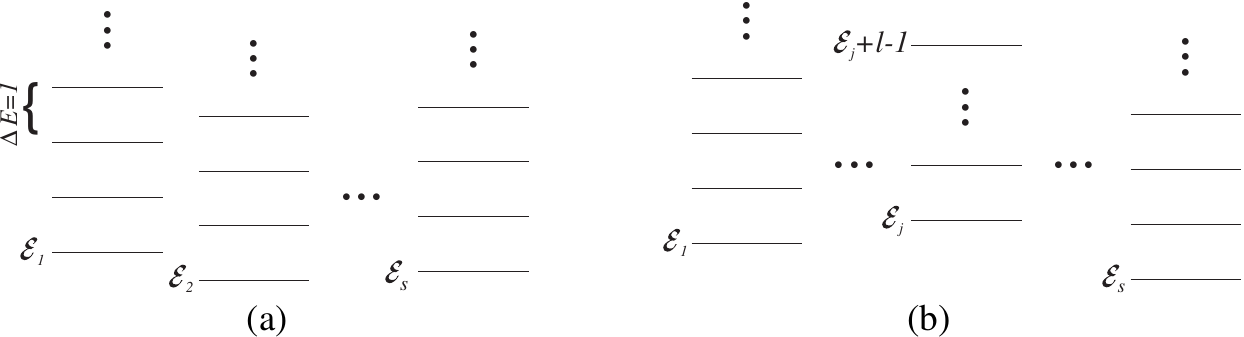}
\caption{Spectrum of the Hamiltonians ruled by polynomial Heisenberg algebras: (a)~it is composed of $s$ inf\/inite ladders starting
from ${\cal E}_j$, $j=1,\dots,s$; (b)~the $j$-th ladder becomes f\/inite since equation~(\ref{finiteladdercondition}) is satisf\/ied.}\label{Fig3}
\end{figure}

Note that Sp($H$) depends on the number of eigenstates of $H$ belonging to the kernel of $L^-$ which have physical meaning. If $s$ of them are physically acceptable and satisfy
\begin{gather*}
L^-\psi_{{\cal E}_i} = 0, \qquad H\psi_{{\cal E}_i} = {\cal E}_i
\psi_{{\cal E}_i}, \qquad i= 1,\dots,s,
\end{gather*}
thus, Sp($H$) turns out to be composed of $s$ independent inf\/inite ladders, each one of them starting from $\psi_{{\cal E}_i}$, $i= 1,\dots,s$ (see Fig.~\ref{Fig3}a).

On the other hand, for the $j$-th ladder which starts from $\psi_{{\cal E}_j}$ it could happen that
\begin{gather}\label{finiteladdercondition}
  (L^+)^{l-1}\psi_{{\cal E}_j} \neq 0, \qquad (L^+)^{l}\psi_{{\cal E}_j} = 0,
\end{gather}
for some integer $l$. In this case it turns out that~\cite{fh99}
\begin{gather*}
{\cal E}_n = {\cal E}_j + l, \qquad n\in\{s+1,\dots,k\},
\end{gather*}
which means that the $j$-th ladder starts from the eigenvalue ${\cal E}_j$ and ends at ${\cal E}_j + l - 1$, i.e., it is a f\/inite ladder of length~$l$, with $l$ steps (see Fig.~\ref{Fig3}b).

\section{Supersymmetric quantum mechanics}\label{section4}

Let us consider now the following chain of intertwining relationships:
\begin{gather}
  H_i A_i^+ = A_i^+ H_{i-1},  \label{intertwining-i-(i-1)} \\
  A_i^\pm = \frac{1}{\sqrt{2}}\left[\mp\frac{d}{d x} +
\alpha_i(x,\epsilon_i)\right], \qquad i = 1,\dots,k , \label{Ai-intertwiners}
\end{gather}
where
\begin{gather}
  H_i = -\frac12 \frac{d^2}{d x^2} + V_i(x), \qquad i=0,\dots,k . \label{i-Hamiltonian}
\end{gather}
By plugging the expressions (\ref{Ai-intertwiners}), (\ref{i-Hamiltonian}) into
equation~(\ref{intertwining-i-(i-1)}), it turns out that the following must be
satisf\/ied:
\begin{gather}
  \alpha_i'(x,\epsilon_i) + \alpha_i^2(x,\epsilon_i) = 2[V_{i-1}(x) - \epsilon_i],
\label{i-rieq} \\
  V_{i}(x) = V_{i-1}(x) - \alpha_i'(x,\epsilon_i).  \label{i-pot}
\end{gather}
Suppose now that $V_{i-1}(x)$ is known; then $V_{i}(x)$ becomes determined (see equation~(\ref{i-pot})) if the solution $\alpha_i(x,\epsilon_i)$ of the $i$-th Riccati equation~(\ref{i-rieq}) associated to $\epsilon_i$ can be found. The key point in this treatment
is to realize that there is a simple f\/inite dif\/ference formula allowing to f\/ind
algebraically $\alpha_i(x,\epsilon_i)$ in terms of two solutions of the $(i-1)$-th
Riccati equation, associated to the factorization energies $\epsilon_{i-1}$, $\epsilon_{i}$ \cite{fhm98,mnr00}:
\begin{gather*}
  \alpha_{i}(x,\epsilon_{i}) = - \alpha_{i-1}(x,\epsilon_{i-1}) -
\frac{2(\epsilon_{i-1}-
\epsilon_{i})}{\alpha_{i-1}(x,\epsilon_{i-1}) -
\alpha_{i-1}(x,\epsilon_{i})} .
\end{gather*}
By iterating down this equation it turns out that, at the end, $\alpha_{i}(x,\epsilon_{i})$
can be expressed in terms of the $i$ solutions
\begin{gather*}
  \alpha_1'(x,\epsilon_j) + \alpha_1^2(x,\epsilon_j) = 2 [V_0(x) - \epsilon_j], \qquad
j=1,\dots,i,
\end{gather*}
of the initial Riccati equation, or in terms of the corresponding solutions of the
Schr\"odinger equation,
\begin{gather}
  - \frac12 u_j'' + V_0(x)u_j = \epsilon_j u_j, \qquad
j=1,\dots,i , \label{ise}
\end{gather}
where $\alpha_1(x,\epsilon_j) = u_j'/u_j$.

In order to connect the previous technique and supersymmetric quantum mechanics,
let us realize now the standard SUSY algebra with two generators
\begin{gather*}
[{\sf Q}_i, {\sf H}_{\rm ss}]=0, \qquad \{{\sf Q}_i,{\sf Q}_j\} =
\delta_{ij} {\sf H}_{\rm ss}, \qquad i,j=1,2,
\end{gather*}
in the following way \cite{ais93,aicd95,bs97,cf08,cjnp08,fe10,fe97,ff05,in04,mr04}
\begin{gather*}
  {\sf Q} = \left(
\begin{matrix}
0 & B_k^+ \\ 0 & 0
\end{matrix}
\right), \qquad
{\sf Q}^+ =  \left(
\begin{matrix}
0 & 0 \\ B_k & 0
\end{matrix}
\right),
\\   {\sf Q}_1 =\frac{{\sf Q}^+ + {\sf Q}}{\sqrt{2}}, \qquad  {\sf Q}_2 =
\frac{{\sf Q}^+ - {\sf Q}}{i\sqrt{2}}, \qquad
   {\sf H}_{\rm ss} = \left(
\begin{matrix}
B_k^+ B_k  & 0 \\ 0 & B_k B_k^+
\end{matrix}
\right),
\end{gather*}
where
\begin{gather}
  B_k^+ B_k = (H_k - \epsilon_1) \cdots (H_k - \epsilon_k),\qquad  
  B_k B_k^+ = (H_0 - \epsilon_1) \cdots (H_0 - \epsilon_k), \label{factorizationk2}
\end{gather}
$H_0$ and $H_k$ being the initial and f\/inal Hamiltonians, intertwined by
$k$-th order dif\/ferential intertwining operators, namely,
\begin{gather}\label{intertwiningk}
  H_k B_k^+ = B_k^+ H_0, \qquad H_0 B_k = B_k H_k,\qquad
  B_k^+ = A_k^+ \cdots A_1^+, \qquad B_k = A_1^-\cdots A_k^-.
\end{gather}
The initial and f\/inal potentials $V_0$, $V_k$, are interrelated by:
\begin{gather*}V_k(x) = V_0(x) - \sum_{i=1}^k\alpha_i' (x,\epsilon_i) = V_0(x) -
\{\ln[W(u_1,\dots,u_k)]\}'',
\end{gather*}
where $W(u_1,\dots,u_k)$ is the Wronskian of the $k$ Schr\"odinger seed solutions $u_1,\dots,u_k$. Let
us note a certain resemblance of the $k$-th order SUSY QM presented here with the method of fractional
supersymmetry discussed elsewhere~\cite{dk06}.

The previous technique has been employed successfully to generate
new solvable poten\-tials~$V_k(x)$ departing from a given initial one~$V_0(x)$ for
several interesting physical systems \cite{ba01,cks01,fe10,mr04}. Of our particular interest is the case of the
harmonic oscillator \cite{bf11b,cfnn04,ff05, fh99,fhm98,fnn04}, which is worth of an explicit discussion.

\section{Harmonic oscillator SUSY partners}\label{section5}

\begin{figure}[t]
\centering
\includegraphics[scale=0.6]{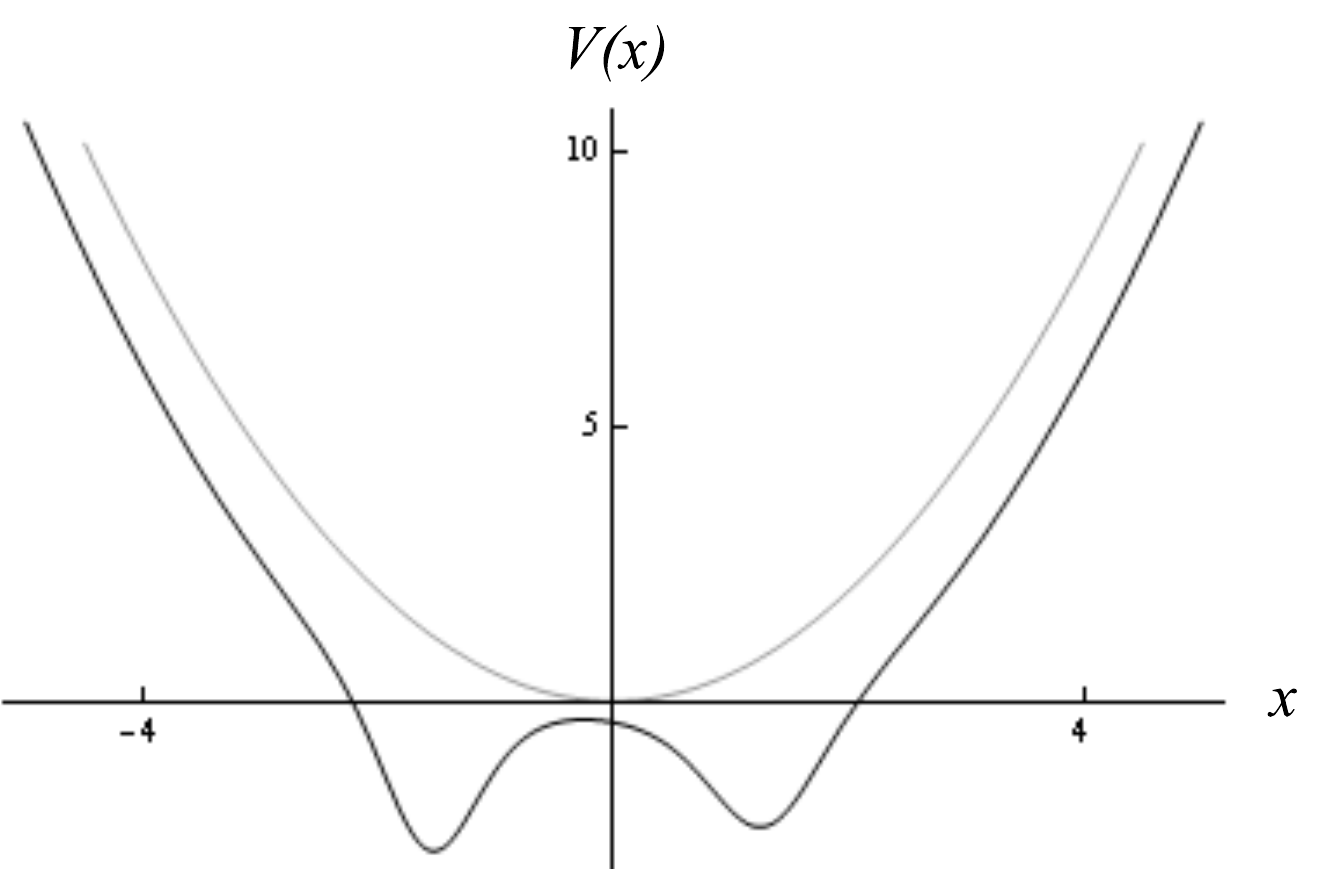}
\caption{Second-order SUSY partner (black curve) of the oscillator potential (gray curve) generated by taking two Schr\"odinger solutions of the form given in equation (\ref{oss}) with $(\epsilon_1,\epsilon_2) = \left(-1,-\frac65\right)$ and $(\nu_1,\nu_2) = (0,2)$.}\label{Fig4}
\end{figure}

In order to implement the SUSY technique, we need to f\/ind f\/irst the general solution
of the stationary Schr\"odinger equation (\ref{ise}) for $V_0(x) = x^2/2$ and an
arbitrary $\epsilon$, which turns out to be:
\begin{gather}
  u = e^{-\frac{x^2}2}\left[ {}_1F_1\left(\frac{1-2\epsilon}{4},\frac12;x^2\right) +
2x\nu\frac{\Gamma(\frac{3 -
2\epsilon}{4})}{\Gamma(\frac{1-2\epsilon}{4})}\,
{}_1F_1\left(\frac{3-2\epsilon}{4},\frac32;x^2\right)\right] , \label{oss}
\end{gather}
where ${}_1F_1(a,b;y)$ is the (Kummer) conf\/luent hypergeometric function.
Let us perform now a~non-singular $k$-th order SUSY transformation which creates precisely $k$ new levels,
by simplicity placed below the ground state energy $E_0 = 1/2$ of the oscillator \cite{fh99}. If the
factorization energies are ordered as $\epsilon_k < \epsilon_{k-1} < \cdots < \epsilon_1 < 1/2$,
then the non-singular SUSY transformations arise for $\vert \nu_1\vert < 1, \vert \nu_2\vert > 1,
\vert \nu_3\vert < 1, \dots$ The spectrum of the corresponding Hamiltonian reads:
\begin{gather}
{\rm Sp}(H_k)  = \{\epsilon_k,\dots,\epsilon_1, \; E_n = n + 1/2,n=0,1,\dots\}. \label{spectrumhk}
\end{gather}
An illustration of a second-order SUSY partner potential of the oscillator for $(\epsilon_1,\epsilon_2) = (-1,-\frac65)$ and $(\nu_1,\nu_2) = (0,2)$ is shown in Fig.~\ref{Fig4}.

\begin{figure}[t]
\centering
\includegraphics[scale=0.55]{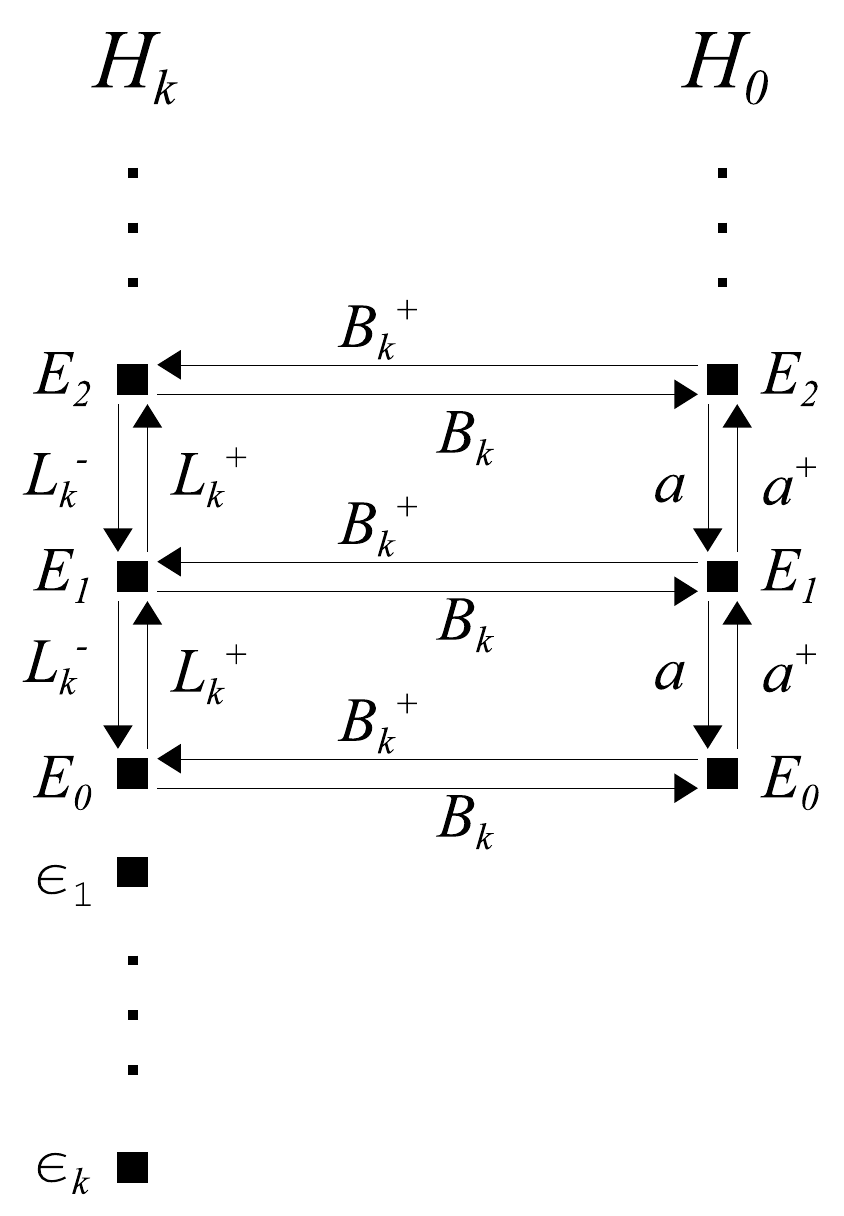}
\caption{Diagram representing the spectrum and the action of the ladder operators $L_k^\pm$
for the SUSY partner~$H_k$ of the harmonic oscillator Hamiltonian~$H_0$. They are
built up from the intertwining opera\-tors~$B_k$,~$B_k^+$ and the harmonic oscillator annihilation and creation
operators $a$, $a^+$ (see equation~\eqref{naturallk}).}\label{Fig5}
\end{figure}

Let us note that, for the Hamiltonian $H_k$, there exists a natural pair of ladder operators:
\begin{gather}
L_k^- = B_k^+ a B_k, \qquad L_k^+ = B_k^+ a^+ B_k, \label{naturallk}
\end{gather}
which are dif\/ferential operators of order $(2k+1)$-th satisfying \cite{cfnn04,fh99, mi84}:
\begin{gather}
[H_k, L_k^\pm] = \pm L_k^\pm. \label{comhklk}
\end{gather}
By making use of the intertwining relationships (\ref{intertwiningk}) and the factorizations of equations 
(\ref{factorizationk2}),
it is straightforward to show that:
\begin{gather*}
Q_{2k+1}(H_k) = L_k^+ L_k^- = \left(H_k - \frac12\right) \prod_{i=1}^k \left(H_k -
\epsilon_i - 1\right) \left(H_k - \epsilon_i \right).
\end{gather*}
This implies that the operators $\{H_k, L_k^-, L_k^+\}$ generate a polynomial Heisenberg algebra
of order $(2k)$-th, which is characterized by equation~(\ref{comhklk}) and the deformed commutator:
\begin{gather*}
[L_k^-,L_k^+]=P_{2k}(H_k).
\end{gather*}
From the analysis of the roots involved in $Q_{2k+1}(H_k)$, it turns out that the ${\rm Sp}(H_k)$ given in equation (\ref{spectrumhk}) can be seen as containing $k+1$ ladders: $k$ of them are one-step ladders, starting and ending at $\epsilon_j$, $j=1,\dots,k$; in addition, there is an inf\/inite one starting from~$\frac12$ \cite{cfnn04, fh99}. A~representation of Sp($H_k$) and the actions of the ladder operators~$L_k^\pm$ are given in Fig.~\ref{Fig5}.

We have already all the elements for answering the main question we would like to pose in this paper: since the harmonic oscillator Hamiltonian has an evolution loop, it is natural to ask if its SUSY partners
show as well such a kind of closed dynamical processes. First of all let us write down the evolution operator $U(t) = e^{-iH_kt}$ associated to $H_k$:
\begin{gather*}
U(t) = \sum_{j=1}^{k} e^{-i\epsilon_j t}\vert \psi_{\epsilon_j}^k \rangle
\langle\psi_{\epsilon_j}^k\vert
+ \sum_{n=0}^{\infty} e^{-i(n+\frac12)t} \vert \psi_{n}^k \rangle\langle\psi_{n}^k\vert .
\end{gather*}
If the $k$ factorization energies $\epsilon_j$, $j=1,\dots,k$ are arbitrary, it turns out that a {\it partial evolution loop} is produced for $\tau = 2\pi$ since:
\begin{gather*}
U(2\pi) = \sum_{j=1}^{k} e^{-i 2\pi\epsilon_j}\vert \psi_{\epsilon_j}^k \rangle
\langle\psi_{\epsilon_j}^k\vert
+ e^{-i \pi} \sum_{n=0}^{\infty} \vert \psi_{n}^k \rangle\langle\psi_{n}^k\vert .
\end{gather*}
This means that any state belonging to the subspace generated by $\{ \vert \psi_{n}^k \rangle, n=0,1,\dots\}$,
\begin{gather*}
\vert\psi(0)\rangle = \sum_{n=0}^{\infty} c_n \vert \psi_{n}^k \rangle, \qquad
\sum_{n=0}^{\infty} \vert c_n\vert^2 = 1,
\end{gather*}
becomes cyclic with period $\tau = 2\pi$:
\begin{gather*}
\vert\psi(\tau)\rangle = e^{-i \pi} \vert\psi(0)\rangle.
\end{gather*}
A straightforward calculation leads now to the associated geometric phase (see equation~(\ref{timeindependentgp})):
\begin{gather}
\beta = -\pi + 2\pi \langle \psi(0) \vert H_k \vert \psi(0) \rangle
= 2\pi \sum_{n=1}^{\infty} n \vert c_n\vert^2. \label{gpsubgeneral}
\end{gather}
In particular, if $c_n = \delta_{n,m}$ it turns out that
\begin{gather*}
\beta = 0 \ \ [{\rm mod}(2\pi)].
\end{gather*}
It is interesting as well to evaluate the geometric phases associated to the coherent states which are
eigenstates of the annihilation operator of equation~(\ref{naturallk})~\cite{fh99}, namely,
\begin{gather*}
  L_k^- \vert z\rangle = z \vert z\rangle, \qquad z\in {\mathbb C}.
\end{gather*}
By expressing $\vert z\rangle$ in terms of the eigenstates of $H_k$,
\begin{gather*}
  \vert z\rangle = \sum_{j=1}^{k} b_j \vert \psi_{\epsilon_j}^k \rangle
+ \sum_{n=0}^{\infty} c_n \vert \psi_{n}^k \rangle ,
\end{gather*}
it turns out that
\begin{gather*}
  b_j = 0, \qquad c_{n} = \frac{z}{\sqrt{n \prod\limits_{i=1}^{k}(n - \epsilon_i - \frac12)
(n - \epsilon_i + \frac12)}} c_{n-1}.
\end{gather*}
By iterating down the recurrence relationship for $c_n$, it turns out that it
becomes expressed in terms of $c_0$. The last coef\/f\/icient is f\/ixed from the normalization
condition, leading to:
\begin{gather*}
  \vert z\rangle = N(r) \sum\limits_{n=0}^{\infty} \frac{z^n \vert \psi_{n}^k \rangle }{\sqrt{n! \prod\limits_{i=1}^{k}
\Gamma(n+\frac12-\epsilon_i)\Gamma(n+\frac32-\epsilon_i)}} , \\
  N(r) = \sqrt{\frac{\prod\limits_{i=1}^{k}\Gamma(\frac12 - \epsilon_i)\Gamma(\frac32 - \epsilon_i)}{
{}_0F_{2k}(\frac12-\epsilon_1,\dots,\frac12-\epsilon_k,\frac32-\epsilon_1,\dots,\frac32-\epsilon_k;r^2)}} ,
\end{gather*}
where $r = \vert z\vert$. By using equation (\ref{gpsubgeneral}), the associated geometric phase becomes now:
\begin{gather}\label{gengpkek}
   \beta = \frac{2\pi r^2}{\prod\limits_{i=1}^{k}(\frac12 - \epsilon_i)(\frac32-\epsilon_i)} \frac{{}_0F_{2k}(\frac32-\epsilon_1,\dots,\frac32-\epsilon_k,\frac52-\epsilon_1,\dots,\frac52-\epsilon_k;
r^2)}{{}_0F_{2k}(\frac12-\epsilon_1,\dots,\frac12-\epsilon_k,\frac32-\epsilon_1,\dots,\frac32-\epsilon_k;r^2)}.
\end{gather}

\begin{figure}[t]
\centering
\includegraphics[scale=0.55]{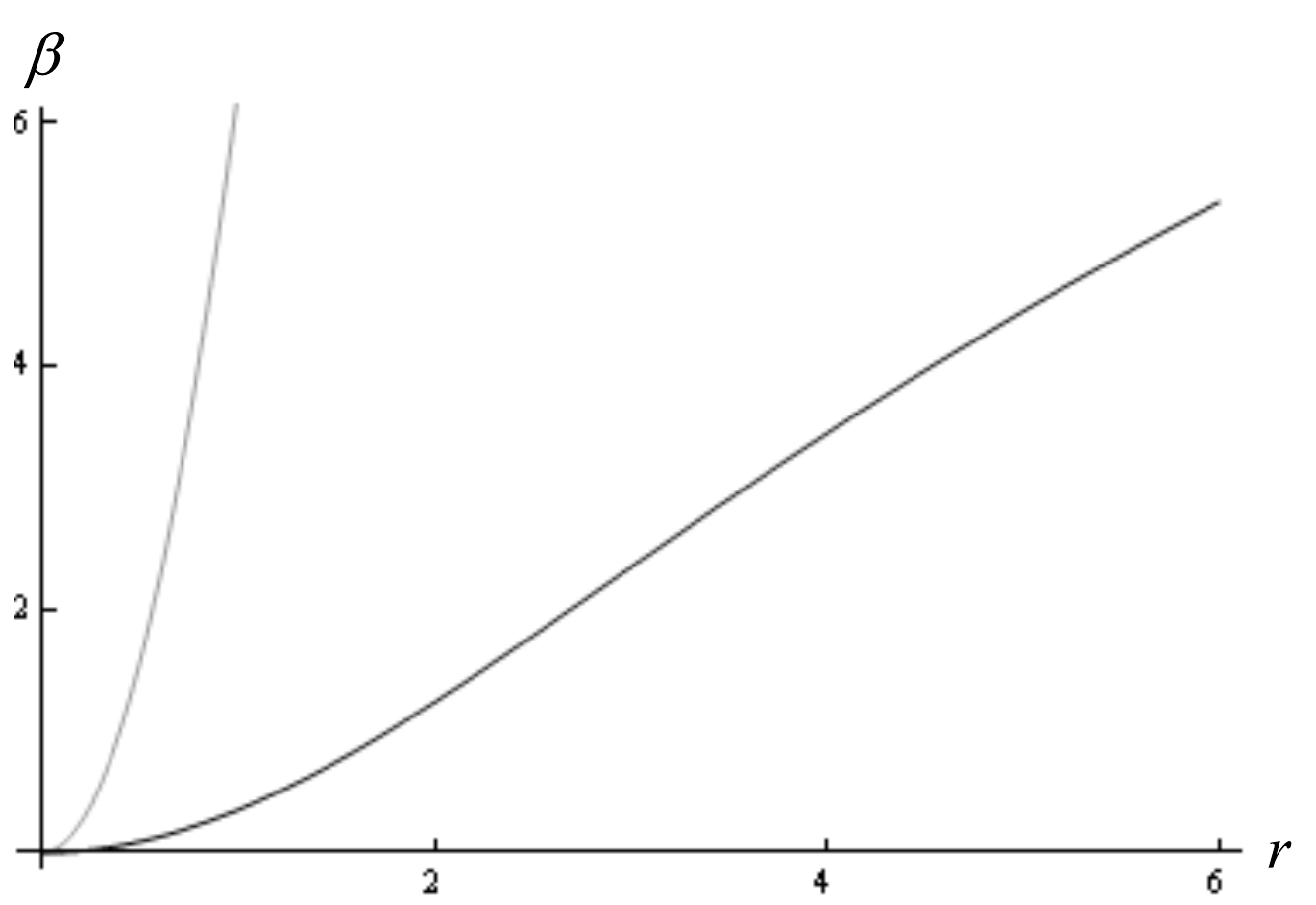}

\caption{Plot of the geometric phase $\beta$ of equation~(\ref{particulargeometricphase}) as a function of~$r$ (black curve). The geometric phase associated to the standard coherent states $\beta = 2\pi r^2$ is as well plotted (gray curve).}\label{Fig6}
\end{figure}

Let us note that, for $k=1$ and $\epsilon_1=-1/2$, the expression of equation (\ref{gengpkek})
reduces (mod\,($2\pi$)) to the expression of equation~(27) of~\cite{fe94} since
\begin{gather*}
\beta_{GCS} - 2\pi = 2\pi\left[ \frac{{}_0F_2(1,1;r^2)}{{}_0F_2(1,2;r^2)} - 1 \right] =
\pi r^2 \, \frac{{}_0F_2(2,3;r^2)}{{}_0F_2(1,2;r^2)}.
\end{gather*}
On the other hand, for $(\epsilon_1,\epsilon_2) = (-1,-\frac65)$ it turns out that
\begin{gather}\label{particulargeometricphase}
\beta = \frac{160\pi r^2}{1377} \frac{{}_0F_{4}(\frac52,\frac72,\frac{27}{10},\frac{37}{10};
r^2)}{{}_0F_{4}(\frac32,\frac52,\frac{17}{10},\frac{27}{10};r^2)} .
\end{gather}
A plot of this geometric phase as a function of $r$ is shown in Fig.~\ref{Fig6} (black curve). The geometric phase
acquired by the standard coherent states, which turns out to be $\beta = 2\pi r^2$, is as well drawn
(gray curve).

Coming back to our general subject, if the factorization energies are such that
\begin{gather*}
\epsilon_j = \frac12 - \frac{l_j}{m_j} \qquad \forall \;j=1,\dots,k,
\end{gather*}
where $l_j$, $m_j$ are coprime, then a {\it global
evolution loop} of period $\tau = 2M\pi$ is obtained, namely,
\begin{gather*}
U(2M\pi)   =   e^{-i M\pi} \left(\sum_{j=1}^{k} e^{i 2\pi l_j\frac{M}{m_j}}\vert \psi_{\epsilon_j}^k \rangle
\langle\psi_{\epsilon_j}^k\vert +  \sum_{n=0}^{\infty} e^{-i 2\pi Mn} \vert \psi_{n}^k \rangle\langle\psi_{n}^k\vert\right) \nonumber \\
\hphantom{U(2M\pi)}{} =   e^{-i M\pi} \left( \sum_{j=1}^{k} \vert \psi_{\epsilon_j}^k \rangle\langle\psi_{\epsilon_j}^k\vert +
\sum_{n=0}^{\infty} \vert \psi_{n}^k \rangle\langle\psi_{n}^k\vert\right) = e^{-i M\pi} I ,
\end{gather*}
with $M$ being the least common multiple of $\{m_j, \;j=1,\dots,k\}$. In this case any arbitrary initial state
\begin{gather*}
  \vert\psi(0)\rangle = \sum_{j=1}^{k} b_j \vert \psi_{\epsilon_j}^k \rangle
+ \sum_{n=0}^{\infty} c_n \vert \psi_{n}^k \rangle, \qquad
\sum_{j=1}^{k} \vert b_j\vert^2 + \sum_{n=0}^{\infty} \vert c_n\vert^2 = 1 ,
\end{gather*}
is cyclic with period $\tau = 2M\pi$:
\begin{gather*}
\vert\psi(\tau)\rangle = e^{-i M\pi} \vert\psi(0)\rangle .
\end{gather*}
The associated geometric phase becomes f\/inally:
\begin{gather*}
   \beta = -M\pi + 2M\pi\!\left( \sum_{j=1}^{k} \epsilon_j \vert b_j\vert^2 + \sum_{n=0}^{\infty} \left( n+\frac12 \right) \vert c_n\vert^2 \right) = 2M\pi\!\left( \sum_{n=1}^{\infty} n \vert c_n\vert^2 -
\sum_{j=1}^{k} \frac{l_j}{m_j} \vert b_j\vert^2
\right).
\end{gather*}

\section{Conclusions}\label{section6}

In this paper it has been shown that the SUSY partners of the harmonic oscillator Hamiltonian realize
straightforwardly the polynomial Heisenberg algebras of order $2k$. As a consequence, if the
SUSY transformation creates $k$ new levels for $H_k$, the corresponding spectrum can be seen as composed
of $k+1$ independent ladders, $k$ of them being one-step ladders and an inf\/inite one starting from
$E_0 = 1/2$. It has been proven also that the corresponding Hamiltonians present in general the so-called
partial evolution loops, which induce cyclic evolutions in the subspace generated by the eigenstates associated
to~$E_n$ and, consequently, have associated geometric phases. This applies, in particular, to the
coherent states which are eigenstates of the natural annihilation operator of the system, for which a general
formula for the associated geometric phase has been derived in this paper. In particular, from this general
result we have recovered the expression for the geometric phase which was found previously~\cite{fe94}
for the Abraham--Moses family of potentials isospectral to the harmonic oscillator. Finally, we have shown that
it is possible to produce global evolution loops by imposing restrictions on the involved factorization energies.
The associated geometric phases induced by the last operator identity have been as well evaluated.

\subsection*{Acknowledgements}

The author acknowledges the f\/inancial support of Conacyt, project 152574, as well as the comments of Alonso Contreras-Astorga.

\pdfbookmark[1]{References}{ref}
\LastPageEnding

\end{document}